\documentclass[reprint,amsmath, superscriptaddress, amssymb, aps, longbibliography, prl,floatfix]{revtex4-1}

\usepackage{graphicx}
\usepackage{dcolumn}
\usepackage{bm}
\usepackage{nccmath}

\usepackage[T1]{fontenc}
\usepackage[utf8]{inputenc}
\usepackage{physics}

\usepackage{xcolor}
\newcommand\hl[1]{%
  \bgroup
  \hskip0pt\color{blue}%
  #1%
  \egroup
}
\DeclareMathAlphabet{\pazocal}{OMS}{zplm}{m}{n}

\begin{document}

\title{Direct visualization of hybrid excitons in van der Waals heterostructures}

\author{Giuseppe Meneghini}
\email{giuseppe.meneghini@physik.uni-marburg.de}
\affiliation{%
 Department of Physics, Philipps University of Marburg, 35037 Marburg, Germany}%

\author{Marcel Reutzel}
\affiliation{%
 I. Physikalisches Institut, Georg-August-Universität Göttingen, Göttingen, Germany}%

\author{Stefan Mathias}
\affiliation{%
 I. Physikalisches Institut, Georg-August-Universität Göttingen, Göttingen, Germany}%
 
\author{Samuel Brem}
\affiliation{%
 Department of Physics, Philipps University of Marburg, 35037 Marburg, Germany}%

\author{Ermin Malic}
\affiliation{%
 Department of Physics, Philipps University of Marburg, 35037 Marburg, Germany}%

\date{\today}

\begin{abstract}
Van der Waals heterostructures show fascinating physics including trapped moire exciton states, anomalous moire exciton transport, generalized Wigner crystals, etc. Bilayers of transition metal dichalcogenides (TMDs) are characterized by long-lived spatially separated interlayer excitons. Provided a strong interlayer tunneling,  hybrid exciton states consisting of interlayer and intralayer excitons can be formed. Here, electrons and/or holes are in a superposition of both layers. Although crucial for optics, dynamics, and transport, hybrid excitons are usually optically inactive and have therefore not been directly observed yet. Based on a microscopic and material-specific theory, we show that time- and angle-resolved photoemission spectroscopy (tr-ARPES) is the ideal technique  to directly visualize these hybrid excitons. Concretely, we predict a characteristic double-peak ARPES signal arising from the hybridized hole in the MoS$_2$ homobilayer. The relative intensity is proportional to the quantum mixture of the two hybrid valence bands at the $\Gamma$ point. Due to the strong hybridization, the peak separation of more than 0.5 eV can be resolved in ARPES experiments. Our study provides a concrete recipe of how to directly visualize hybrid excitons and how to distinguish them from the usually observed regular excitonic signatures. 
\end{abstract}

\maketitle

\begin{figure}[t!]
  \centering
  \includegraphics[width=\columnwidth]{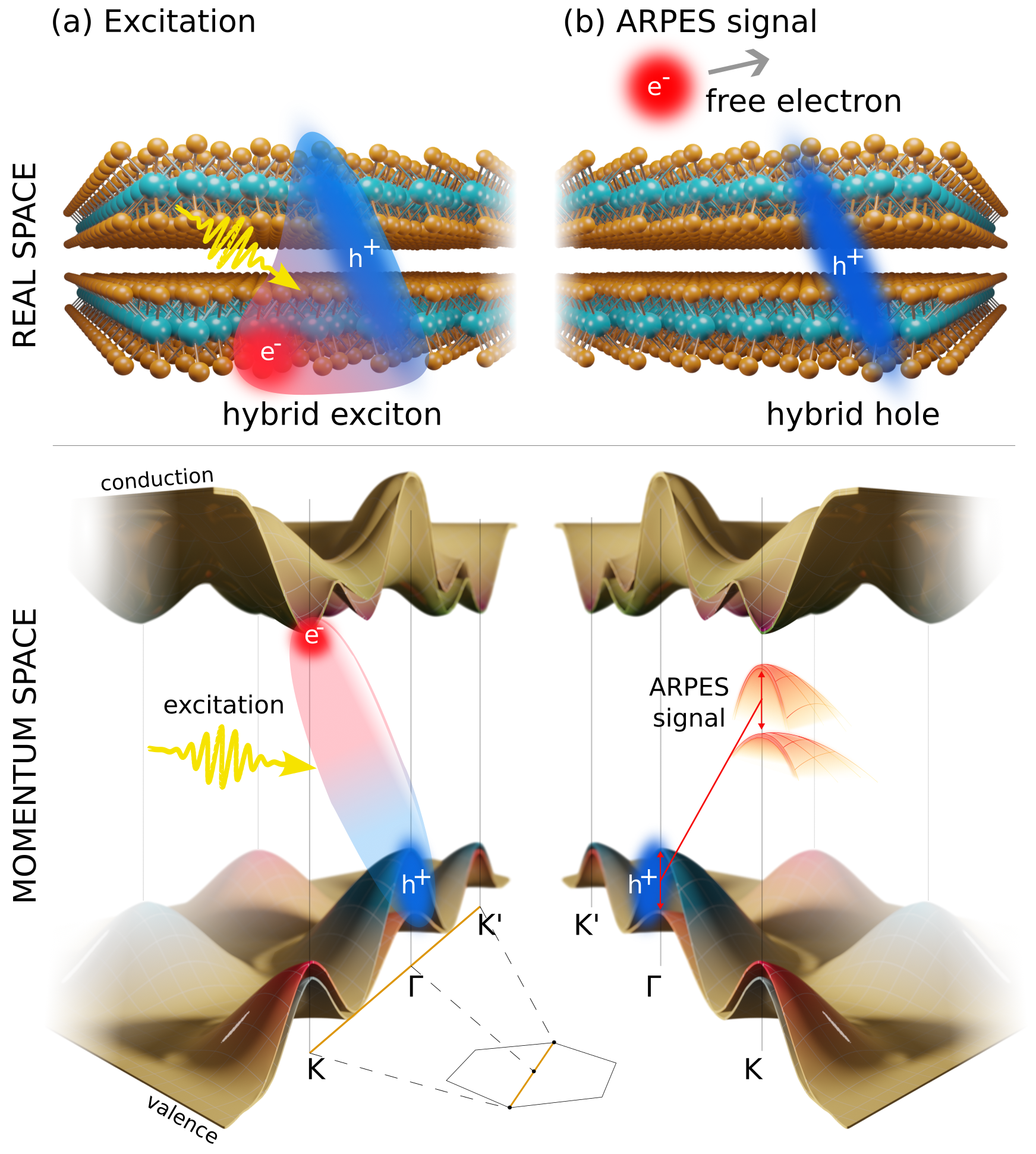}
  \caption{ARPES signature of hybrid excitons. The figure represents the system in real and momentum space (a) before and (b) after a photon-induced electron ejection from an excited MoS$_2$ homobilayer. Here, the most energetically favorable state is a momentum-dark $\Gamma$K hybrid exciton with the hole being strongly delocalized between the layers. The hybrid exciton breaks down into an ejected free electron and a superposition of hybrid holes. This gives rise to a characteristic double-peak ARPES signal reflecting the hole superposition between the two hybrid valence bands at the $\Gamma$ point. The signal appears at the K point and exhibits the negative curvature of the $\Gamma$ valence bands.}
   \label{fig:scheme}
\end{figure}

The research on atomically thin nanomaterials has become one of the most active fields in condensed matter physics, showing fascinating phenomena ranging from moire exciton effects to exotic strongly correlated states. Here, the material class of transition metal dichalcogenides (TMDs) has been in the focus of many investigations due to their unprecedented properties. TMD monolayers are characterized by tightly bound excitons that govern optics, dynamics and transport phenomena at room temperature \cite{he2014tightly, chernikov2014exciton, wang2018colloquium,mueller2018exciton}. Artificially stacked TMD bilayers exhibit long-lived spatially separated interlayer excitons, where the Coulomb-bound electrons and holes are located in different layers. Furthermore, due to a large tunneling probability (in particular in TMD homobilayers), hybrid excitons (hX) appear,  in which Coulomb-bound electrons and/or holes are strongly delocalized over the two layers, cf. Fig. \ref{fig:scheme}(a). These new quasi-particles can be considered as a quantum superposition of the involved electron and hole states in both layers.

The importance of interlayer excitonic states has been demonstrated in a series of experiments \cite{rivera2015observation,miller2017long,kunstmann2018momentum, jin2019observation,tran2019evidence,seyler2019signatures,alexeev2019resonantly,ruiz2019interlayer, sigl2022optical, holler2022interlayer}. They exhibit a long lifetime and a large dipole moment and thus offer a possibility of controlling exciton optics and transport by external electric fields \cite{ciarrocchi2019polarization,sun2022excitonic,hagel2022electrical,thygesen18}. In contrast hybrid excitons have been less intensively studied, although they combine a high oscillator strength (intralayer excitons) with a sensitivity to electric fields (interlayer excitons) enabling tunability of their properties. It is meanwhile well known that hX are important in particular in TMD homobilayers \cite{brem2020hybridized,hagel2021exciton}, however, they have not been directly observed, yet. 
Recently, time- and angle-resolved photoemission spectroscopy (tr-ARPES) has been demonstrated as a powerful technique to directly visualize momentum-dark exciton states in TMD monolayers as well as interlayer exciton dynamics in TMD heterostructures \cite{schmitt2022formation,wallauer2021momentum,christiansen2019theory,madeo2020directly,dong2021direct}.Here, we show that this technique can be further exploited to even map out the wavefunction of hybrid exciton states. 
Based on a fully microscopic and material-specific approach we model the tr-ARPES signal in TMD bilayers. We focus on the exemplary material system of a MoS$_2$ homobilayer in H$^h_h$ stacking exhibiting a large hybridization of excitons \cite{hagel2021exciton}. We predict the emergence of a characteristic double-peak ARPES signal reflecting the strongly hybridized hole at the $\Gamma$ point that is left behind after the ejection of the electron. Here, the relative intensity of the peaks sensitively depends on the degree of hole hybridization. This is in strong contrast to the single excitonic signal observed so far in ARPES experiments. The developed method has been applied to the exemplary TMD homobilayer, but it is of general nature and thus applicable to a wide range of exciton-dominated material systems. 
\bigskip

\noindent
\textit{\bf ARPES signal of excitons:}
To microscopically model the hX dynamics, we first start by defining the electronic Hamilton operator for a TMD bilayer, including tunnelling between the two layers. The required material-specific input parameters are obtained from first-principle calculations \cite{kormanyos2015k,ferreira2021band}. Then, we transform the Hamiltonian first into an exciton and then into a hybrid exciton basis. This allows us to determine the hybrid exciton band structure and calculate the optical selection rules by solving the Wannier equation for a bilayer system \cite{hagel2021exciton,brem2020tunable,brem2020hybridized, merkl2020twist}. Furthermore, we obtain microscopic access to the hybrid exciton relaxation dynamics via exciton-phonon scattering, where we describe the thermalization process after optical excitation via the Boltzmann scattering equation \cite{meneghini2022ultrafast}. These are the key ingredients for a material-specific model for ARPES experiments. Here, it is crucial to define the initial and final states, as illustrated in Fig. \ref{fig:scheme}. First, the TMD bilayer is optically excited using a laser pulse. The subsequent relaxation into the energetically most favourable state creates a population of hybrid excitons (Fig. \ref{fig:scheme}(a)). In the second step, the system is illuminated with a second laser pulse that breaks the Coulomb-bound electron-hole pairs into ejected electrons and remaining holes (Fig. \ref{fig:scheme}(b)). In an ARPES measurement, the ejected electron is  collected in a detector, leaving behind the hole in the material. The latter - in the case of hybrid excitons - will remain in a superposition of the two valence band states that where involved in the photo-dissociated exciton (cf. the split valence bands at the $\Gamma$ point in Fig. \ref{fig:scheme}(b)).
The ARPES signal allows to reconstruct the exciton energy from the electron information measured, since the Coulomb-bound electrons and holes present a strongly correlated system.

The photoemission probability can be described using a time‐dependent perturbation theory yielding Fermi's golden rule \cite{damascelli2003angle} for the ARPES signal
\begin{align}
    \mathcal{I}({\bf k},h\nu; t) \propto \sum_{i f} \,\lvert \bra{f, \bf k} H_{int} \ket{i}\rvert^2 N_i(t)\,
    \delta\left(\Delta E_{f,i, \bf k} \right)
  \label{eq:arpes1}
\end{align}
where $\Delta E_{f,i, \bf k}=\hbar^2k^2/(2m_0)+E_{f}- E_i - h\nu$ with the the free-electron momentum ${\bf k}$ and the photon energy  $h\nu$. Furthermore, $\ket{i/f}$ are the initial/final states of the system  with eigenergies $E_{i/f}$ and the initial state occupation $N_i(t)$. The interaction Hamiltonian    $H_{int} = \sum_{\alpha\beta} \mathcal{M}_{\alpha\beta} a^{\dagger}_{\mathrm{f}\alpha} a^{}_{c\beta}$ describes the excitation of an electron from the conduction band to the free state with $a^{(\dagger)}$ denoting the electronic creation/annihilation operator. Note that we use the suffix $\mathrm{f}$ for a free state and $c/v$ for a conduction/valence band state. The optical selection rules are contained in the optical-matrix element $\mathcal{M}_{\alpha \beta}$.

Since the interaction Hamiltonian is defined in terms of single-particle operators, it is important to also determine the initial  and final states $\ket{i}$, $\ket{f}$ entering this equation in a single-particle basis.
To reach this, we perform a series of transformations to express the hybrid exciton states with electron operators, as detailed in the supplementary material. 
The final state  is described by an uncorrelated product of a free electron state and a hybrid hole state, that we derive by solving the eigenvalue problem for the electronic Hamiltonian of a bilayer system, obtaining hybridized valence and conduction bands $E^\lambda_{{\bf k}\gamma}$ with $\gamma = (\pm,\xi_\lambda)$ being the compound index consisting of the quantum number for the split layer-hybridized bands ($\pm$), the valley index $\xi = \Gamma,\Lambda^{(\prime)},K^{(\prime)}$, and the band index  $\lambda = c,v$. 
We denote the splitting of the hybrid conduction or valence bands with $\Delta E^\lambda_{\bf k} = E^\lambda_{{\bf k}+}-E^\lambda_{{\bf k}-}$, i.e. in particular the hybrid valence band splitting at the K and $\Gamma$ point relevant for this work read $\Delta E^v_{K}$ and $\Delta E^v_{\Gamma}$, respectively. 

Inserting the initial and final state discussed above (and explicitly shown in the SI) in Eq. (\ref{eq:arpes1}) we obtain the final equation for the ARPES signal of hybrid excitonic states 
\begin{equation}\label{eq:arpes2}
    \mathcal{I}({{\bf k}},h\nu; t) \propto \sum_{\eta \gamma {{\bf p}} }\, \lvert  \mathcal{G}^{\eta\gamma }_{{\bf p} {\bf k}}\rvert^2 \, N^\eta_{{\bf k} -{\bf p}}(t)  \,\delta\left(\Delta E^\eta_{\gamma,  \bf p, \bf k}\right)
\end{equation}
where $\Delta E^\eta_{\gamma,  \bf p, \bf k}=E^{e}_{\bf k} - E^{v}_{\bf \gamma,p } - E^{X}_{\eta,{\bf k} -{\bf p}} - h\nu$. The ARPES signal is significantly influenced by the hybrid exciton occupation $N^\eta_{\bf Q}(t)$ in the hX state $\eta$ at the center-of-mass momentum ${\bf Q}$. The dynamic occupation is determined microscopically by solving a semiconductor Bloch equation in second-order Born-Markov approximation and explicitly including all phonon-mediated scattering channels within the hybrid exciton landscape \cite{meneghini2022ultrafast}, cf. the SI for more details.
 Moreover, the energy conservation during the photoemission is of key importance with $E^{e}_{\bf k}$ denoting  the free-electron energy, $E^{v}_{\bf p \gamma}$ the hybrid valence band energy, and $E^{X\eta}_{{\bf k} -{\bf p}}$ the hybrid exciton energy. The new optical-matrix element $\mathcal{G}^{\eta \gamma }_{{\bf p} {\bf k}}$ depends on the excitonic wave function and on the overlap between the layer-mixed hybrid hole and hybrid exciton. A detailed derivation is shown in the SI. 

\begin{figure}[t!]
  \centering
  \includegraphics[width=\columnwidth]{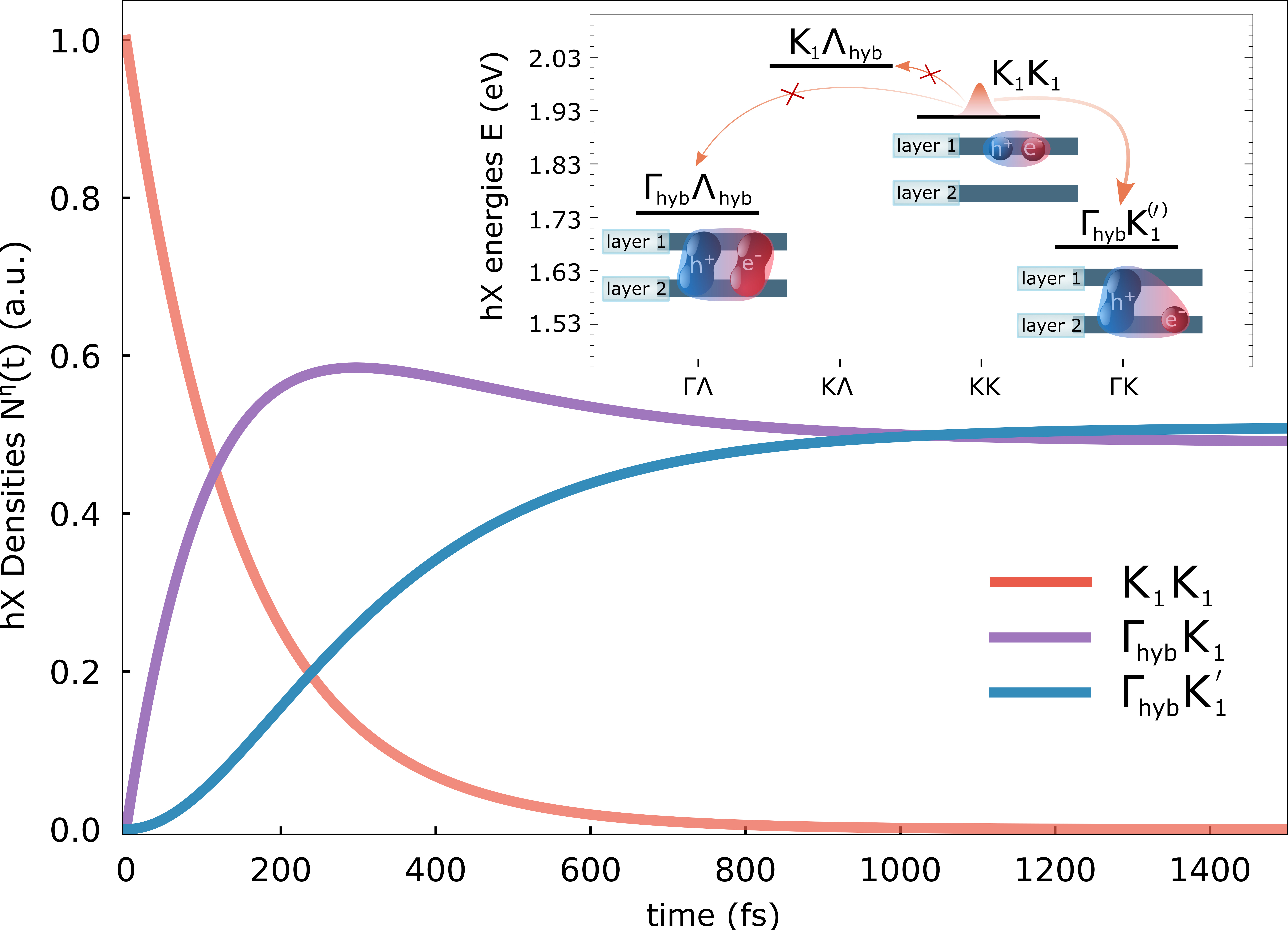}
  \caption{Hybrid exciton dynamics. After optical excitation of a MoS$_2$ homobilayer at $\simeq 1.9 eV$ (resonant to the K$_1$K$_1$ exciton), ultrafast exciton relaxation dynamics occurs resulting in the highest occupation N$^\eta$ of the  energetically most favorable momentum-dark $\Gamma$K$^{(\prime)}$ hybrid exciton (red and purple lines). The inset shows the hybrid exciton dispersion illustrating possible relaxation channels (note that $\Gamma_{hyb}$K$^{(\prime)}$ states are almost degenerate in energy).}
   \label{fig:landscape+dynamics}
\end{figure}

\begin{figure*}[t!]
  \centering
  \includegraphics[width=\textwidth]{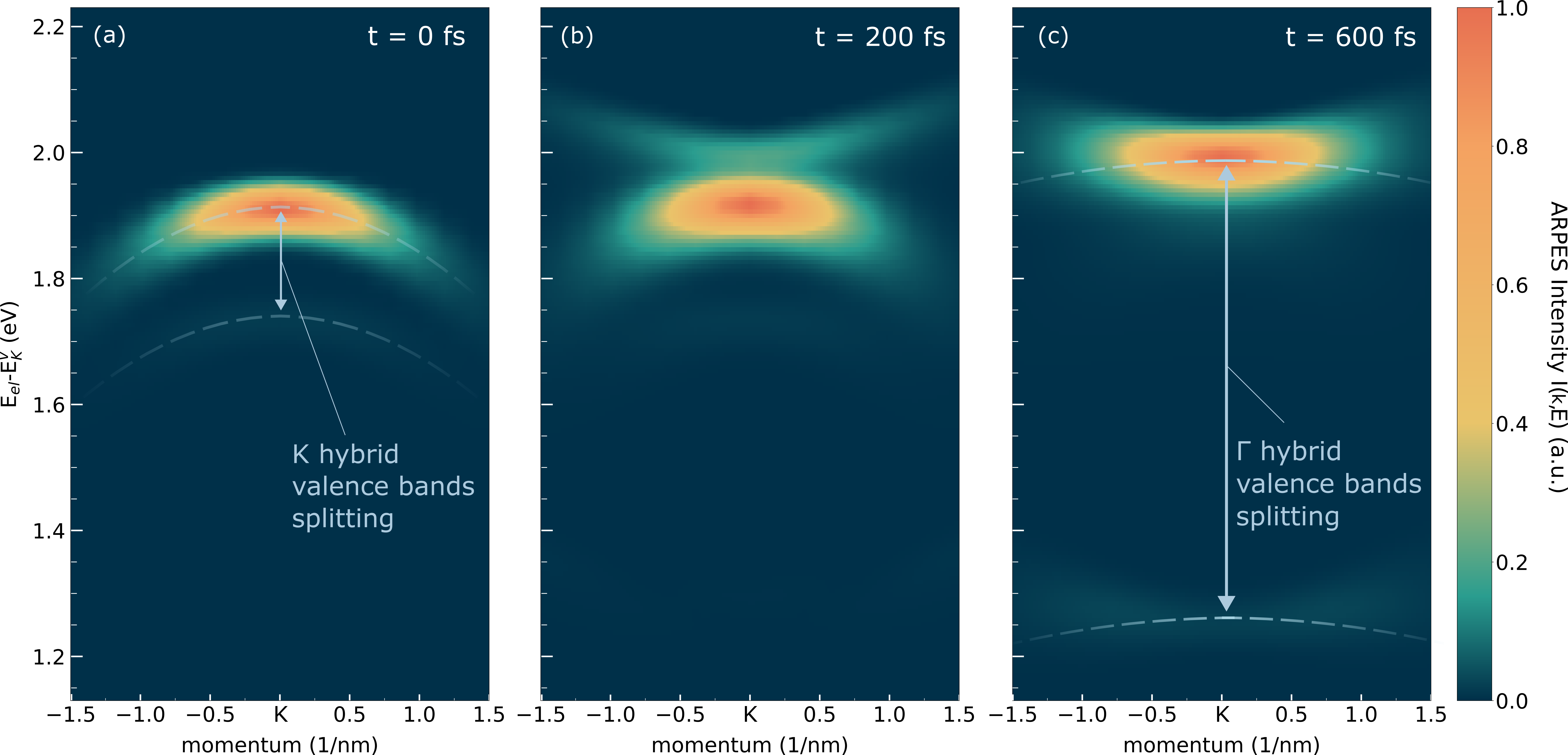}
  \caption{Momentum-resolved ARPES map. Hybrid exciton dynamics tracked in a tr-ARPES signal (a) revealing the excitation of the nearly purely intralayer K$_1$K$_1$ exciton state (0 fs), (b) the formation of the  strongly hybridized $\Gamma_{hyb}$K$_1$ state (100 fs) and (c) the thermalized hybrid exciton distribution (400 fs). The dashed lines show the shifted split valence bands of the hybrid hole at the K point (with the splitting $\Delta E_K^v$) and at the $\Gamma$ point (with the splitting $\Delta E_\Gamma^v$). The energies are shown with respect to the upper hybrid valence band ($E^v_{K}$) at the K point (cf. also Fig. S2 in the SI).}
   \label{fig:Kpath_E}
\end{figure*}

\bigskip

\textit{\bf Hybrid exciton dynamics:}
We exploit the theoretical framework described above to study the exemplary MoS$_2$ homobilayer in naturally available H$^h_h$ stacking. This material is ideal for our study for two reasons: (i) The most energetically favorable states are momentum-dark  $\Gamma_{hyb}$K$^{(\prime)}_1$ hybrid excitons. Due to a lack of lower-lying states and their momentum/layer indirect character, they exhibit long lifetimes facilitating their observation in ARPES spectra.
(ii) The strong interlayer tunneling results in a large splitting of the valance bands at the $\Gamma$ point.
As a result, the hole in $\Gamma_{hyb}$K excitons is delocalized over a large spectral range and we expect to find pronounced hybrid hole signatures in ARPES spectra.

We start with discussing the relaxation dynamics of hybrid excitons, as the temporal evolution of the exciton occupation is of crucial importance for ARPES spectra, cf. Eq. \ref{eq:arpes2}.
Solving the Wannier equation for a bilayer system allows us to resolve the hybrid exciton landscape. We show the relevant states contributing in the relaxation process in the inset of Fig. \ref{fig:landscape+dynamics}.  
For studying the dynamics, we start with an initial exciton population in layer 1 centered around the energy of $\simeq 1.9$ eV, reflecting an optical excitation resonant to the intralayer K$_1$K$_1$ exciton. We solve the equation of motion for the hybrid exciton occupation including all phonon-mediated scattering channels in the low-density regime. This allows us to track the way of initially excited hybrid excitons in momentum and time. Figure \ref{fig:landscape+dynamics} illustrates the relaxation dynamics of momentum-integrated hybrid exciton densities $N^\eta(t)$. We observe an ultrafast  population transfer from the initially occupied K$_1$K$_1$ exciton (red line) - that is almost completely intralayer-like - to the most energetically favorable momentum-dark hybrid exciton states $\Gamma_{hyb}$K$^{(\prime)}_1$ (purple and blue lines). This is followed by a thermalization process in which the charge is redistributed between the two almost degenerate $\Gamma_{hyb}$K$^{(\prime)}_1$ states. The K$\Lambda_{hyb}$ states are not involved in the relaxation process, since they are located above the excitation energy, cf. the inset of Fig. \ref{fig:landscape+dynamics}. Furthermore, the  $\Gamma_{hyb}\Lambda_{hyb}$ exciton could, in principle, be important for the dynamics considering its low energy, however this can be neglected for two main reasons. (i) The direct scattering would require a simultaneous scattering of both electron and hole from the K$_1$K$_1$ state. This two-phonon process is thus of higher-order and can be neglected. (ii) The indirect scattering involving one-phonon processes, K$_1$K$_1$ $\rightarrow$ K$\Lambda_{hyb} \rightarrow \Gamma_{hyb}\Lambda_{hyb}$ and K$_1$K$_1$ $\rightarrow$ $\Gamma_{hyb}$K$^{(\prime)}_1 \rightarrow \Gamma_{hyb}\Lambda_{hyb}$ involve phonon absorption processes and will therefore have a negligible role in the relaxation dynamics.

\bigskip

\textit{\bf Hybrid exciton signatures in ARPES:}
Having determined the hybrid exciton occupation, we can now evaluate Eq. (\ref{eq:arpes2}) to investigate the time- and momentum-resolved ARPES signal in MoS$_2$ homobilayers. In recent studies, it has been shown that the excitonic ARPES signal appears at the momentum corresponding to the electron valley and it is spectrally located one excitonic energy above the valence band (or one exciton binding energy below the conduction band) \cite{weinelt2004dynamics,madeo2020directly,schmitt2022formation}. The shape of the signal is expected to be characterized by the negative curvature of the valence band (where the hole left behind is located), if the exciton population is very sharp in momentum \cite{christiansen2019theory,rustagi2018photoemission}.  Note, however, that for a thermally distributed exciton occupation, the ARPES signal will be smeared out in energy and momentum. 
Figure \ref{fig:Kpath_E} shows the momentum-resolved ARPES map for different time snapshots, where we have fixed the highest valence hybrid band at the K point ($E^v_K$) as a reference energy determining the position of ARPES signals, cf. also Fig. S2 in the SI.
At the center of the optical excitation pulse (0 fs), we observe an ARPES signal reflecting the nature of the almost purely intralayer K$_1$K$_1$ exciton which is characterized by a well pronounced single peak (Fig. \ref{fig:Kpath_E}(a)). We find that on a sub-100 fs timescale, strongly hybridized  $\Gamma_{hyb}$K$^{(\prime)}_1$ excitons are formed. They are characterized by two peaks, one slightly above the K$_1$K$_1$ exciton  and one red-shifted by more than 600 meV (Fig. \ref{fig:Kpath_E}(b)). The last step of the dynamics leads to a thermalization of the hX occupation. After 400 fs, the entire population has reached an equilibrium distribution (Fig. \ref{fig:Kpath_E}(c)), where only the signatures stemming from  $\Gamma_{hyb}$K$^{(\prime)}_1$ excitons have remained.

Regarding the shape of the ARPES signals, we find that a narrow distribution in momentum leads to a negative dispersion reflecting the curvature of the valence band. This can be observed for the initial occupation of intralayer K$_1$K$_1$ excitons in layer 1 (Fig.\ref{fig:Kpath_E} (a)).
The phonon-driven exciton relaxation dynamics and the subsequent thermalization brings the system into a thermal equilibrium. At room temperature, this results in a broad exciton distribution over the center-of-mass momentum. For this reason, the shape of the final ARPES signal is reflecting a mixture of the hybrid valence bands (curved downwards) and the hybrid exciton parabola (curved upwards), cf. Fig.\ref{fig:Kpath_E} (c). A more detailed discussion on how the center-of-mass distribution affects the shape of the signal can be found in the SI.

\begin{figure*}[t!]
  \centering
  \includegraphics[width=\textwidth]{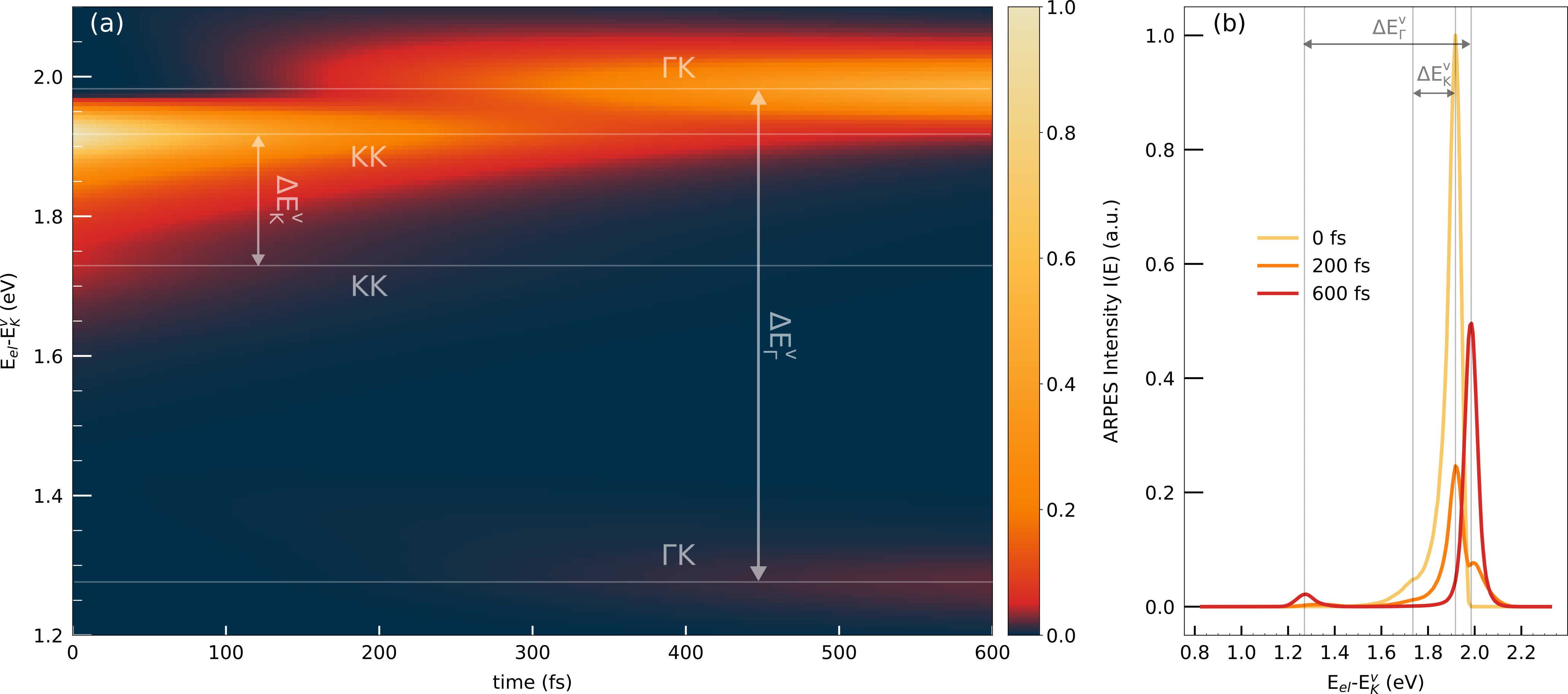}
  \caption{Momentum-integrated ARPES map.  (a) Energy- and time-resolved ARPES signal, showing the characteristic double-peak structure reflecting the hybrid hole being spread over two valence bands at K point (initial signal) and at the $\Gamma$ point (final thermalized signal). (b) ARPES signal at fixed times plotted as a function of energy.  The energies are shown with respect to the upper hybrid valence band ($E^v_{K}$) at the K point as reference.}
   \label{fig:Et}
\end{figure*} 

The most important message of our work is the prediction of a double-peak ARPES signal that is characteristic for hybrid exciton states. For the considered MoS$_2$ homobilayer, the ARPES signal is governed by the energetically lowest $\Gamma_{hyb}$K$_1$ hybrid exciton. 
Figure \ref{fig:Kpath_E}(c) clearly exhibits two peaks that are separated by approx. 0.6 eV. Their position corresponds to the energy of the split hybrid valence bands (illustrated by dashed lines).
To explain this observation, we consider the spatial (layer) distribution of the single-particle states contributing to the entangled electron-hole pair. The two valence bands ($\ket{\pm}$) at $\Gamma$ are completely delocalized across both layers, i.e. $\ket{\pm}=\left(\ket{1}\pm \ket{2}\right)/\sqrt{2}$, where $\ket{n}$ indicates the valence band of the pure monolayer $n$. Now, the hybrid exciton is formed with an electron that is strongly localized in one of the two layers (at the K point), e.g. layer 1. The Coulomb interaction partially drags the hole into the same layer to reduce the energy, favouring a hole wave function that is mostly in layer 1 too, i.e. $\ket{1}=(\ket{+}+\ket{-})/\sqrt{2}$. Consequently, the energetically most favourable two-body state (hX) is build by a superposition of the two hybrid valence bands $\ket{\pm}$.     

When the K$_1$ electron from the $\Gamma_{hyb}$K$_1$ hybrid exciton is ejected, a $\Gamma$ hole is left behind, remaining in the superposition that has previously formed the exciton. 
The conservation of energy and momentum ensures that measuring the energy of the ejected electron, we obtain information about the energy of the hole as well. While the hole within the hybrid exciton is in a quantum mixture of two energy levels, the relative energy between electron and hole is fixed by the two-particle exciton energy $E^{X\eta}$. This entanglement between electron and hole transfers the superposition of hole energies to the correlated electron, whose energy is consequently undefined as well. Measuring the energy of ejected (initially entangled) electrons therefore allows us the reconstruct the underlying energy distribution of the holes.

To resolve this better, Fig. \ref{fig:Et} shows the momentum-integrated ARPES signal. At first glance we observe that the $\Gamma_{hyb}$K$_1$ hybrid exciton has a long lifetime. The ARPES signal remains over picoseconds,  since this exciton is the energetically lowest state without any scattering partners at lower energies (cf. the inset of Fig. \ref{fig:landscape+dynamics}). We find that a clear transfer from the initially excited K$_1$K$_1$ excitons to the momentum-dark $\Gamma_{hyb}$K$_1$ hybrid excitons occurs on a sub-100 fs timescale. In this time, a double-peaked ARPES signal is formed that is characteristic for an hybrid exciton state and that reflects the splitting $ \Delta E^v_ K$ of the hybrid valence bands at the $\Gamma$ point.  Since the single particle states at the K point are strongly layer-polarized, we do not expect to see a well pronounced double-peak ARPES signal for K$_1$K$_1$ excitons.  However, even this state is weakly hybridized and has a small contribution of an interlayer exciton due to a weak tunnelling of holes at the K point. This gives rise to a second signal with low-intensity that is red-shifted by about 200 meV reflecting the splitting of the hybrid valence bands $ \Delta E^v_ K$ at the K point, cf. Fig. \ref{fig:Et}(b). 

In summary, our work demonstrates how hybrid excitons can be  identified in tr-ARPES spectra. The investigated MoS$_2$ homobilayer is an exemplary case and the developed approach can be applied to a much larger class of exciton-dominated materials. 

\bigskip

\textit{\bf Discussion:}
Based on a microscopic and material-specific theory, we predict well pronounced signatures of hybrid exciton in tr-ARPES spectra. 
We investigate the exemplary case of a MoS$_2$ homobilayer in the naturally available H$^h_h$ stacking and find a characteristic double-peak ARPES signal arising from the split hybrid valence bands at the $\Gamma$ point where the left-behind hole is located. This double-peaked signal can be considered as a clear fingerprint for the existence of hybrid exciton states. In particular, materials with an efficient interlayer tunneling resulting in a large spectral splitting of the hybrid valence bands are favorable, as the double-peak signal is then easy to resolve in the experiment. Furthermore, the  presence of energetically lowest dark hybrid exciton states is of advantage as they exhibit a long lifetime facilitating the experimental observation. 
Although the choice of MoS$_2$ homobilayer is favorable for these reasons, the hybridization of the hole (50\% - 50\%) is similar to the exciton hybridization giving rise to the predicted relatively small double-peak intensity ratio. Heterostructures exhibiting a considerably different degree of hybridization for holes and excitons should give rise to larger intensity ratios that are easier accessible in ARPES experiments. 
Overall, our work provides a concrete recipe of how to directly visualize hybrid exciton states in ARPES measurements and will trigger new experimental studies in atomically thin semiconductors and related materials. \\

  \textbf{Acknowledgements:} This project has received funding from Deutsche Forschungsgemeinschaft via CRC 1083 and the project 512604469 as well as from the European Unions Horizon 2020 research and innovation programme under grant agreement no. 881603 (Graphene Flagship).

\bibliography{references}

\end{document}